# Imaging Fractal Conductance Fluctuations and Scarred Wave Functions in a Quantum Billiard

R. Crook, C. G. Smith, A. C. Graham, I. Farrer, H. E. Beere, and D. A. Ritchie
Department of Physics, University of Cambridge, Cambridge CB3 0HE, UK

**We present scanning-probe images and magnetic-field plots which reveal fractal conductance fluctuations in a quantum billiard. The quantum billiard is drawn and tuned using erasable electrostatic lithography, where the scanning probe draws patterns of surface charge in the same environment used for measurements. A periodicity in magnetic field, which is observed in both the images and plots, suggests the presence of classical orbits. Subsequent high-pass filtered high-resolution images resemble the predicted probability density of scarred wave functions, which describe the classical orbits.**

Universal conductance fluctuations, on the scale of $e^2/h$, are observed in disordered systems due to multiple-path interference as electrons scatter from random impurities [1]. A quantum billiard is a large quantum dot where electron trajectories are ballistic, meaning scattering occurs predominantly at the billiard boundary. If the electron phase coherence length is longer than the billiard dimensions, then conductance fluctuations can also be observed in quantum billiards where electrons scatter off the billiard boundary instead of impurities [2-5]. A soft-walled quantum billiard is a classically mixed system, with regions of regular and chaotic behavior, characterized by the presence of fractal magnetoconductance fluctuations [4,6,7]. The system is chaotic in the sense that a small change, in the magnetic field for example, strongly modifies conductance on an arbitrarily fine scale. Quantum billiards often exhibit Aharonov-Bohm like [1] periodic conductance fluctuations, which are understood to be the signature of stable closed-loop orbits with well defined areas whose quantum states are preferentially excited due to collimation from the leads [2]. The amplitude of the associated wave functions, which are known as scarred wave functions, are concentrated along the underlying classical trajectories and are found through simulation to also exist periodically in magnetic field [8-10]. In this letter we provide a further link between experiment and simulation by presenting high resolution scanning probe images of fractal conductance fluctuations which reveal structure remarkably similar to that seen in theoretical images of scarred wave functions [8].

Figure 1 illustrates the billiard construction. A 2D electron system (2DES) with electron mobility $5 \times 10^6$ cm$^2$ V$^{-1}$ s$^{-1}$ and density $3.1 \times 10^{11}$ cm$^{-2}$ forms at a GaAs/AlGaAs heterojunction 97 nm beneath the surface. The billiard is defined from the 2DES using erasable electrostatic lithography (EEL) where a conductive scanning probe draws spots of negative charge on the GaAs surface to locally deplete 2DES electrons [11]. Uniquely, EEL uses the same low-temperature high-vacuum environment as used for measurement, so device geometry can be modified during the experiment. A row of EEL spots, separated by 100 nm, creates a linear barrier in the 2DES which defines the quantum billiard walls. The lithographic dimension of the billiard is 2 by 3.5 μm, but EEL line width and lateral depletion decrease the 2DES billiard dimension to approximately 1.4 by 2.9 μm. Increased EEL charge density in the billiard corners, and inherent material disorder, mean the confining potential is not a regular rectangle, and the billiard is chaotic. Additional EEL charge spots, separated by 10 nm, tune the billiard's leads so that each transmit one degenerate 1D subband ($n = 2$) which is known to maximize fractal behavior [3]. Tuning was monitored by studying conductance plateau as a function of probe bias with the probe positioned over either lead [11]. Throughout all the experiments, a constant $-1$ V bias was applied to the gold surface electrodes shown in Fig. 1, which locally depletes 2DES electrons to isolate the source and drain 2DES regions. A narrow 2DES region forms between the electrodes whose resistance contributes to the 1 k$\Omega$ series resistance subtracted from all the data.

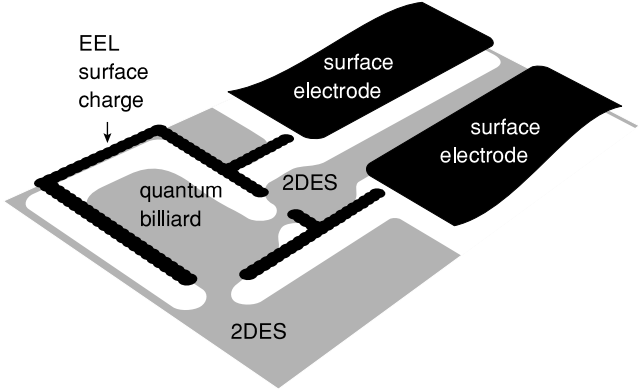

**FIG. 1.** Illustration of the device. Erasable electrostatic lithography charge-spots are drawn on the device surface to define a 1.4 by 2.9 µm quantum billiard from the subsurface 2DES. Biased surface electrodes separate the source and drain 2DES regions.

Fractals are quantified by their fractal dimension $D_F$ which is the scaling relationship between magnification and irregularity [12,13]; curves have $D_F$ between 1 and 2, while surfaces have $D_F$ between 2 and 3. Box-counting is an established technique to calculate $D_F$ where a curve is plotted over a grid of squares called boxes. For magnetoconductance plots, the box has width $\Delta B = X_B / n$ and height $\Delta G = X_G / n$ where $n$ is an integer, $X_B$ is the range of $B$, and $X_G$ the range of $G$. As a function of $n$, the number of occupied squares $N(\Delta B)$ are counted, and a straight line relationship between $\log N(\Delta B)$ and $\log \Delta B$ indicates fractal behavior where the gradient provides $D_F$. Figure 2(a) plots billiard conductance against perpendicular magnetic field. The conductance fluctuations, which are reproducible, are shown to be fractal over the range $1 < \Delta B < 30$ mT by the straight line relationship in Fig. 2(b) which has gradient $D_F = 1.44$. This is close to the theoretical maximum near $D_F = 1.5$ [3,4]. The fluctuations are not fractal for $\Delta B < 1$ mT because $B$ approaches the correlation field, and for $\Delta B > 30$ mT due to a reduced accuracy of the box-counting algorithm for a small number of squares. The low-field half-maximum correlation field is $B_c = 0.3$ mT, and the phase-decoherence time is $\tau_q = 100$ ps calculated by analyzing $B_c$ as a function of magnetic field [14,15]. The phase-decoherence length is $v_F \tau_q = 24$ µm so electrons maintain coherence within the billiard even after multiple boundary scattering events. After subtracting a low-order polynomial fit from the magnetoconductance data [2,9,14], Fig. 2(c) plots the low-field power spectrum revealing a strong oscillation with frequency $f = 0.33$ mT$^{-1}$ (period 3.0 mT) and a weaker oscillation with $f = 0.71$ mT$^{-1}$ (period 1.4 mT). Structure in the spectrum is concentrated below $f = 1$ mT$^{-1}$, which corresponds to an Aharonov-Bohm area $\phi_0 f$ [16] close to the billiard area. This suggests that the oscillations are the signature of well defined closed-loop orbits with associated scarred wave functions [2,8-10]. The magnetic period is smaller than reported elsewhere because period scales inversely with billiard dimension [17], and here the quantum billiard is larger to facilitate imaging.

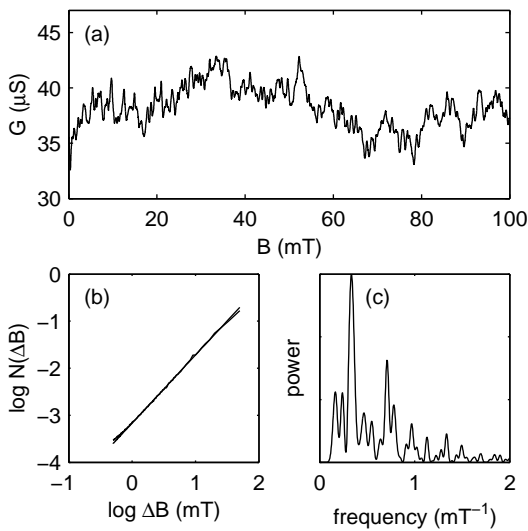

**FIG. 2.** (a) Magnetoconductance of the quantum billiard at 20 mK. (b) Result of box counting algorithm to determine fractal dimension of magnetoconductance data where $D_F = 1.44$. (c) Power spectrum of magnetoconductance data from 0 to 20 mT.

Scanned gate microscopy (SGM) generates images by scanning a biased probe over a quantum device, such as a quantum wire [18-21] or a quantum dot [22-23]. As the probe scans, the device conductance is recorded to set the color of the associated image pixel. Figure 3(a) presents a series of SGM images generated by scanning the probe over the 2 by 2 μm region of the top half of the quantum billiard. The same probe draws EEL patterns and generates SGM images, but crucially, when imaging the tip scans 100 nm off the device surface so the EEL charge patterns are not disrupted. EEL is ideal for this experiment because the absence of topographic surface features mean the probe can scan over the entire billiard area without the possibility of a collision. With the tip biased to 0V, the difference between the tip (silicon) and device (GaAs) surface potentials causes a small local fluctuation in the electron density sufficient to locally perturb the electron wavelength, but insufficient to modify the lead transmission or change the shape of the confining potential. This is confirmed by the presence of oscillations in image correlation with the same periods in magnetic field obtained earlier from the magnetoconductance data. Figure 3(b) plots image correlation against magnetic field period and, in addition to a high correlation between adjacent images, a strong peak is seen at 3.0 mT with a weaker peak at 1.4 mT. Images in Fig. 3(a) are offset vertically to highlight the image correlation horizontally. Many features can be tracked between images, such as the spot in the upper left corner which diminishes between 0 mT and 0.8 mT, and the spot in the bottom right corner which moves upwards then diminishes between 1.8 mT and 2.8 mT.

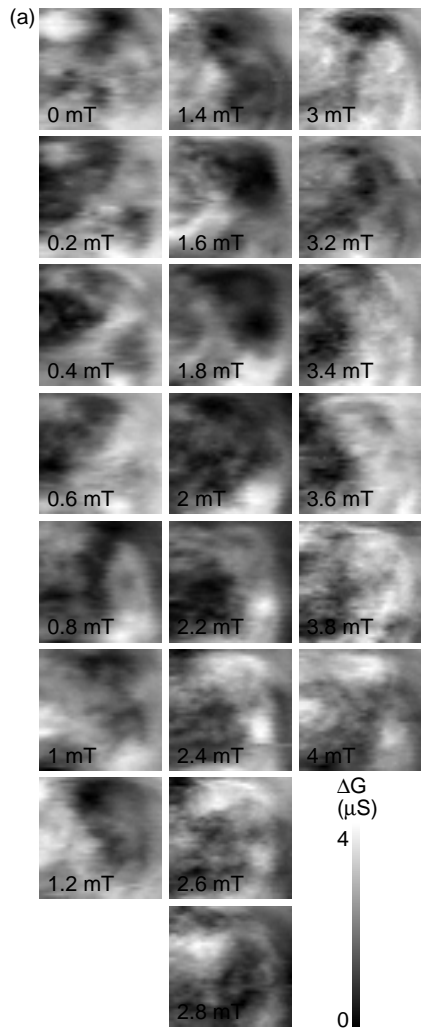

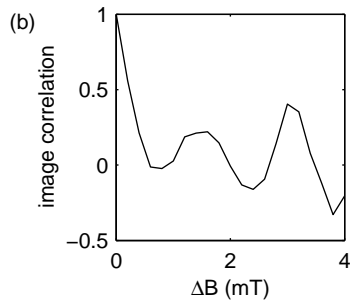

**FIG. 3.** (a) Series of 2 by 2 μm SGM conductance images made by scanning the probe over the top half of the quantum billiards with 0.2 mT increments in magnetic field. Only the top half is scanned for efficiency. (b) Plot of average image correlation as a function of magnetic period.

Motivated by the prospect of imaging scarred wave functions, Fig. 4(a,b) presents high resolution SGM images which reveal a wealth of detail over a range of length scales. The images cover the entire quantum billiard and were made in magnetic fields of 42.0 and 39.1 mT respectively. At these higher magnetic fields a strong periodicity in magnetoconductance is seen at 7.0 mT as shown by the peak at 0.14 mT$^{-1}$ in the power spectrum of Fig. 4 (d). This periodicity is also observed in the images as a correlation of $-0.29$ between images (a) and (b) which are separated by about half this magnetic period. The magnetic periodicity suggests the continued presence of scarred wave functions but associated with a different family of orbits.

Broad structure dominates the SGM images, so high-pass filtering is used to identify features associated with electron interference such as scarred wave functions. The procedure is analogous to subtracting a low-order polynomial from the magnetoconductance data before Fourier analysis. Figure 4(c) presents the high-pass filtered version of image (b), and the filtered image reveals general features resembling scarred wave functions [8]. To highlight the resemblance, three types of structure are identified by annotation in Fig. 4(f) on a copy of image (c). First, the boldest features appear to be enclosed within a boundary which approximately traces the perimeter of the EEL defined billiard which is shown on all the images. Boundary distortion is understood to be caused by inherent device disorder or imperfect EEL patterning, and rounding caused by increased EEL surface charge density in the corners. Second, lines are identified in the vertical direction, parallel to the billiard long axis. The plot across image (f) is an average of profiles across the width of the 1.4 µm central region of the billiard, and peaks identify the vertical lines. Lines parallel to flat boundaries are a common feature of scarred wave functions, being associated with stable orbits scattering between or around boundaries [8]. Third, two small regions, or nodes, where the images look blurred are identified by image (e) which is an image of the local deviation of image (c). Similar features are often seen in scarred wave functions where electron trajectories spend less time. With the additional observation that the lines and nodes reflect the symmetry of the billiard boundary, this is convincing evidence that general features of electron orbits, described by scarred wave functions, have been imaged.

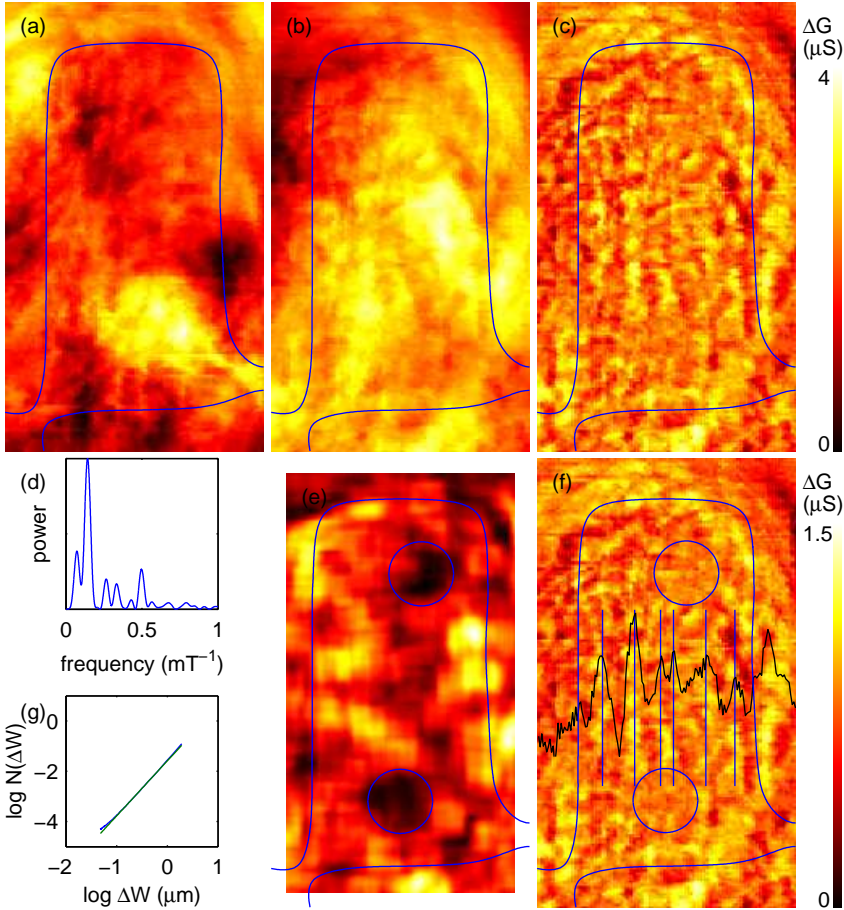

**FIG. 4.** (a,b) High resolution 2 by 3.5 µm SGM conductance images of the quantum billiard in magnetic fields of 42.0 and 39.1 mT. The approximate billiard boundary is shown. (c) High-pass filtered version of image (b). (d) Power spectrum of magnetoconductance data from 30 to 50 mT. (e) Deviation map of image (c) highlighting nodes. (f) Copy of image (c) with profile plot and annotation to identify features characteristic of scarred wave functions. (g) Result of cube-counting algorithm used to determine the fractal dimension of image (b) where $D_F = 2.19$.

To further the discussion of the images, note the following points. The measurement is only sensitive to the probe modulation to interference of closed-loop orbits, and therefore does not detect the direct trajectory between the leads. Energy calculations predict that the billiard leads are coupled to many billiard wave functions, some of which are scarred wave functions. Therefore, image (c) reveals general features of several scarred wave functions, and not specific features from a single scarred wave function. Even the smallest imaged features are several pixels wide and therefore several electron wavelengths long (pixel width $= 25$ nm, $\lambda_F = 45$ nm), so only the boldest features of scarred wave functions are resolved, and fine electron-interference detail is not. High-pass filtering of images at other magnetic fields reveals similar features, although the features are more pronounced at certain fields, which is consistent with the interpretation as scarred wave functions.

To investigate the fractal nature of the SGM images the box-counting algorithm is extended to cube-counting. $N(\Delta W)$ is the number of occupied cubes with width $\Delta W = X_W/n$, height $2\Delta W$, and depth $\Delta G = X_G/n$ where $X_W$ is the image width. The images are fractal over length scales $0.1 < \Delta W < 1$ μm with fractal dimensions $D_F = 2.23$ and 2.19. The images are not fractal for $\Delta W < 0.1$ μm because $\Delta W$ approaches the electron wavelength where the structure resembling wave functions is known to exist, and for $\Delta W > 1$ μm due to a reduced accuracy in the cube-counting algorithm for small numbers of cubes. Figure 4(g) shows the result of the cube counting algorithm for SGM image (b). Note that the filtered images are not fractal. Given that the system is chaotic, it is not surprising that the SGM images are fractal as well as the magnetoconductance data, although the significance of the image fractal dimension, which varies by a few percent with $B$, is not fully understood. The fractal nature of the images accounts for the large range of length scales seen in the structure of Figs. 3 and 4.

Erasable electrostatic lithography has been used to fabricate and tune a quantum billiard to exhibit fractal conductance fluctuations with a high fractal dimension. Subsequent high-pass-filtered scanned-gate-microscopy images strongly resemble theoretical scarred wave functions. These experiments demonstrate the power of combining erasable electrostatic lithography with scanned probe microscopy to tune geometry and image coherent quantum phenomena. Future experiments will likely focus on smaller billiards where a one-to-one correspondence would be expected with simulation. Scanning probe measurements of quantum dots are expected to image electron interference or electron probability density patterns with a resolution of the electron wavelength.

We thank B. D. Simons, C. H. W. Barnes, and C. J. B. Ford for discussions, and S. J. O'Shea for scanning probe design. This work was supported by the EPSRC (UK).